\documentclass[twoside,twocolumn,9pt]{article}
\usepackage{extsizes}
\usepackage[super,sort&compress,comma]{natbib}
\usepackage[version=3]{mhchem}
\usepackage[left=1.5cm, right=1.5cm, top=1.785cm, bottom=2.0cm]{geometry}
\usepackage{balance}
\usepackage{mathptmx}
\usepackage{sectsty}
\usepackage{graphicx}
\usepackage{lastpage} 
\usepackage[format=plain,justification=justified, singlelinecheck=false,font={stretch=1.125,small,sf},labelfont=bf,labelsep=space]{caption} 
\usepackage{float}
\usepackage{fancyhdr}
\usepackage{fnpos}
\usepackage[english]{babel}

\usepackage{array}
\usepackage{droidsans}
\usepackage{charter}
\usepackage[T1]{fontenc}
\usepackage[usenames,dvipsnames]{xcolor}
\usepackage{setspace}
\usepackage[compact]{titlesec}
\usepackage{hyperref}
\usepackage{physics}
\usepackage{nicefrac}
\usepackage{subcaption}
\usepackage{chemformula}
\usepackage{xspace}
\usepackage{nicefrac}
\usepackage{ulem} 

\usepackage{amsmath}
\usepackage{amssymb}
\usepackage{amscd}
\usepackage[np,autolanguage]{numprint}
\usepackage{bm}

\usepackage{xcolor}

\usepackage{hyperref}
\usepackage{cleveref}    
\usepackage{braket}      
\usepackage{upgreek}     
\usepackage{chemformula} 
\usepackage{booktabs}    
\definecolor{cream}{RGB}{222,217,201}

\begin{document}

\pagestyle{fancy}
\thispagestyle{plain}
\fancypagestyle{plain}{
\renewcommand{\headrulewidth}{0pt}
}

\makeFNbottom
\makeatletter
\renewcommand\LARGE{\@setfontsize\LARGE{15pt}{17}}
\renewcommand\Large{\@setfontsize\Large{12pt}{14}}
\renewcommand\large{\@setfontsize\large{10pt}{12}}
\renewcommand\footnotesize{\@setfontsize\footnotesize{7pt}{10}}
\makeatother

\renewcommand{\thefootnote}{\fnsymbol{footnote}}
\renewcommand\footnoterule{\vspace*{1pt}%
\color{cream}\hrule width 3.5in height 0.4pt \color{black}\vspace*{5pt}}
\setcounter{secnumdepth}{5}

\makeatletter
\renewcommand\@biblabel[1]{#1}
\renewcommand\@makefntext[1]%
{\noindent\makebox[0pt][r]{\@thefnmark\,}#1}
\makeatother
\renewcommand{\figurename}{\small{Fig.}~}
\sectionfont{\sffamily\Large}
\subsectionfont{\normalsize}
\subsubsectionfont{\bf}
\setstretch{1.125} 
\setlength{\skip\footins}{0.8cm}
\setlength{\footnotesep}{0.25cm}
\setlength{\jot}{10pt}
\titlespacing*{\section}{0pt}{4pt}{4pt}
\titlespacing*{\subsection}{0pt}{15pt}{1pt}

\fancyfoot{}
\fancyfoot[LO,RE]{\vspace{-7.1pt}\includegraphics[height=9pt]{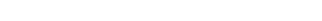}}
\fancyfoot[CO]{\vspace{-7.1pt}\hspace{11.9cm}\includegraphics{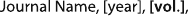}}
\fancyfoot[CE]{\vspace{-7.2pt}\hspace{-13.2cm}\includegraphics{head_foot/RF}}
\fancyfoot[RO]{\footnotesize{\sffamily{1--\pageref{LastPage} ~\textbar  \hspace{2pt}\thepage}}
}
\fancyfoot[LE]{\footnotesize{\sffamily{\thepage~\textbar\hspace{4.65cm} 1--\pageref{LastPage}}
}}
\fancyhead{}
\renewcommand{\headrulewidth}{0pt}
\renewcommand{\footrulewidth}{0pt}
\setlength{\arrayrulewidth}{1pt}
\setlength{\columnsep}{6.5mm}
\setlength\bibsep{1pt}

\makeatletter
\newlength{\figrulesep}
\setlength{\figrulesep}{0.5\textfloatsep}

\newcommand{\topfigrule}{\vspace*{-1pt}%
\noindent{\color{cream}\rule[-\figrulesep]{\columnwidth}{1.5pt}} }

\newcommand{\botfigrule}{\vspace*{-2pt}%
\noindent{\color{cream}\rule[\figrulesep]{\columnwidth}{1.5pt}} }

\newcommand{\dblfigrule}{\vspace*{-1pt}%
\noindent{\color{cream}\rule[-\figrulesep]{\textwidth}{1.5pt}} }

\makeatother

\twocolumn[
  \begin{@twocolumnfalse}
{\includegraphics[height=30pt]{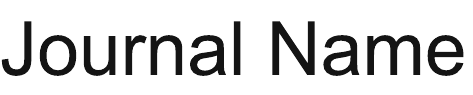}\hfill\raisebox{0pt}[0pt][0pt]{\includegraphics
[height=55pt]{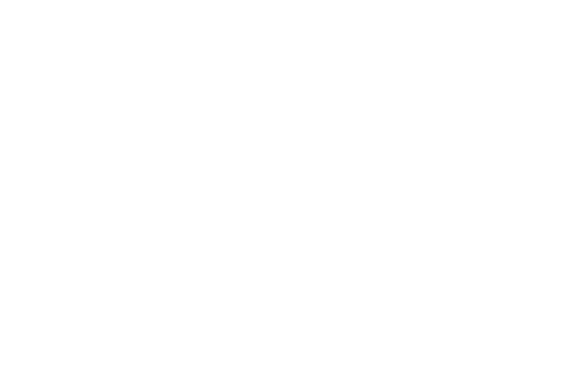}}\\[1ex]
\includegraphics[width=18.5cm]{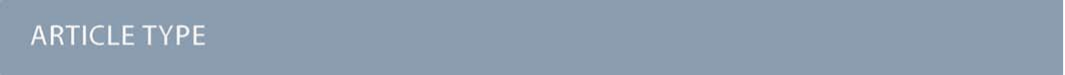}}\par
\vspace{1em}
\sffamily
\begin{tabular}{m{4.5cm} p{13.5cm} }

\includegraphics{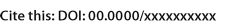} & \noindent\LARGE{\textbf{The effects of dispersion damping and three-body interactions for accurate layered-material exfoliation energies$^\dag$}} \\
\vspace{0.3cm} & \vspace{0.3cm} \\

 & \noindent\large{Adrian F. Rumson,\textit{$^{a}$} Kyle R. Bryenton,\textit{$^{a}$} and Erin R.\ Johnson\textit{$^{a,b,c\ast}$}} \\

\includegraphics{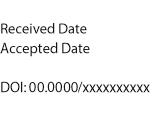} & \noindent\normalsize{%
Accurate predictions of exfoliation energies and lattice constants of layered materials hinge on a correct description of London dispersion physics. Modern \textit{a posteriori} dispersion corrections in density-functional theory (DFT), such as the exchange-hole dipole moment (XDM) model, capture the proper asymptotic behaviour at long range while making use of damping functions to prevent unphysical divergence at short range. In the united-atom limit, the dispersion energy is damped to a finite, non-zero value by both the canonical Becke--Johnson (BJ) damping function and the new Z-damping function.
XDM(BJ) has previously demonstrated exceptional accuracy for modelling layered materials, such as in the LM26 benchmark, which includes graphite, hexagonal boron nitride, lead(II) oxide, and transition-metal dichalcogenides. This work presents the first assessment of XDM(Z) on the same benchmark. We also show that inclusion of three-body interactions via the Axilrod--Teller--Muto (ATM) term further improves the computed exfoliation energies for both XDM(BJ) and XDM(Z), yielding the best performance achieved on LM26 using semi-local functionals to date, relative to reference data from the random-phase approximation. 
} \\

\end{tabular}

 \end{@twocolumnfalse} \vspace{0.6cm}

  ]

\renewcommand*\rmdefault{bch}\normalfont\upshape
\rmfamily
\section*{}
\vspace{-1cm}


\footnotetext{\textit{$^{a}$~Department of Chemistry, Dalhousie University, 6243 Alumni Crescent, Halifax, Nova Scotia, B3H 4R2, Canada. E-mail: erin.johnson@dal.ca}}

\footnotetext{\textit{$^{b}$~Department of Physics \& Atmospheric Science, Dalhousie University, 6274 Coburg Rd, Halifax, Nova Scotia, B3H 4R2, Canada.}}

\footnotetext{\textit{$^{c}$~Yusuf Hamied Department of Chemistry, University of Cambridge, Lensfield Road, Cambridge, CB2 1EW, United Kingdom.}}

\footnotetext{\dag~Electronic Supplementary Information (ESI) available, see DOI: 10.1039/cXCP00000x/}



\section{Introduction}

Layered materials, such as graphite, hexagonal boron nitride, lead(II) oxide, and transition-metal dichalcogenides (TMDCs), are distinguished by their crystal structures, which consist of stacks of atomically thin layers. Because the interlayer binding is due to the fairly weak, London dispersion interactions, the layers can be separated by shearing or direct application of force. The resulting stable monolayers hold exceptional promise as 2D semiconductors,\cite{chhowalla2016two,choi2017recent,manzeli20172d,wang2019two,gupta2020comprehensive,rumson2025role} optoelectronics,\cite{bonaccorso2010graphene,mahmoudi2018graphene,an2022perspectives} and sensors.\cite{he2012graphene,anichini2018chemical} Bulk layered materials can support the intercalation of atoms for energy-storage applications,\cite{li201830,li2021alkali} and facile interlayer sliding makes them excellent solid-state lubricants.\cite{savage1948graphite,lee2010frictional,berman2018approaches}

Density-functional theory (DFT) is a nearly ubiquitous method for the computational study of layered materials, and description of London dispersion is essential for accurate prediction of interlayer binding. Some density functionals, such as the non-local van der Waals density functional (vdW-DF) family\cite{dion2004van,roman2009efficient,lee2010higher} and (r)VV10,\cite{vydrov2010nonlocal,sabatini2013nonlocal} include dispersion directly. Alternatively, a geometry-dependent or post-self-consistent dispersion correction may be applied to add these interactions to functionals that do not otherwise include dispersion physics. Well-known examples include the Grimme-D,\cite{grimme2004accurate, grimme2006semiempirical, grimme2010consistent, grimme2011effect, caldeweyher2019generally} many-body dispersion (MBD),\cite{tkatchenko2009accurate, tkatchenko2012accurate, ambrosetti2014long, hermann2020density, kim2020umbd, gould2016fractionally} or exchange-hole dipole moment (XDM)\cite{xdmchapter, johnson2006post, otero2012van, bryenton2026consistent} corrections.

A key quantity used in benchmarking DFT methods for layered materials is the exfoliation energy,  $E_\text{exfol}$, which is the energy penalty associated with isolating a monolayer from the bulk. It is commonly expressed as an energy per unit area to facilitate comparison between different layered materials:
\begin{equation}
    E_{\text{exfol}} = -\frac{1}{A}\left(\frac{E_{\text{bulk}}}{N_\text{layers}} - E_{\text{mono}}\right)  \,.
\end{equation}
Here, $E_\text{bulk}$ and $E_\text{mono}$ are the energies of the bulk material and an isolated monolayer, respectively, $A$ is the in-plane area of the unit cell, and $N_\text{layers}$ is the number of layers in the cell. Further, a potential energy curve can be mapped in terms of the unit-cell $c$-parameter, corresponding to the lattice vector perpendicular to the atomic layers, which determines the interlayer spacing. 

The best available reference data for exfoliation energies comes from random-phase approximation (RPA) calculations for 26 layered materials performed by Bj\"orkmann,\cite{bjorkman2014testing} which we will refer to as the ``LM26'' benchmark.
The RPA is expected to provide an accurate description of long-range dispersion interactions, but has long been known to give an inadequate description of correlation at short range\cite{singwi1968electron,kurth1999density} and slightly underestimates $C_6$ dispersion coefficients.\cite{gould2012communication} Both Ning \textit{et al.}\cite{ning2022workhorse} and Krogel \textit{et al.}\cite{krogel2020perspectives} note that RPA underestimates exfoliation energies for graphite, \ch{MoS2}, and \ch{TiS2} compared to experiment and Quantum Monte Carlo data by 5--10~meV/\AA$^2$. Similar behaviour is noted by Zen \textit{et al.}, comparing high-level reference methods for molecular crystals.\cite{zen2018fast} However, in the absence of higher-level reference data for a set as large as LM26, comparisons to RPA data can still offer insight and have been common in the literature.\cite{bjorkman2014testing,tawfik2018evaluation,otero2020asymptotic,ning2022workhorse,emrem2022london}

It was previously shown that vdW-DFs are typically accurate for computing exfoliation energies, but overestimate interlayer spacings compared to the RPA.\cite{bjorkman2014testing} Conversely, for a subset of the LM26 (which we refer to as LM11), Tawfik \textit{et al.}\cite{tawfik2018evaluation} found that combinations of generalised gradient approximation (GGA) functionals and pairwise dispersion corrections provided accurate geometries, but overestimated exfoliation energies. Their conclusion was that, to achieve both acceptably accurate geometries and exfoliation energies simultaneously, one must use either use the fractionally ionic many-body dispersion method (FIA-MBD) or the SCAN meta-GGA functional\cite{sun2015strongly} paired with the non-local rVV10 dispersion method.\cite{vydrov2010nonlocal,sabatini2013nonlocal}  Afterwards, Ning \textit{et al.}\cite{ning2022workhorse} confirmed that SCAN-rVV10 performs strongly for the full LM26 set, plus two additional systems: \ch{NbS2} and \ch{VTe2}. However, Otero-de-la-Roza \textit{et al.}\cite{otero2020asymptotic} showed that combining GGA functionals with the XDM dispersion correction gives comparable performance at a lower computational cost, due to the more sophisticated environment dependence of the XDM dispersion coefficients compared to those of other pairwise methods. 

\begin{figure}
    \centering
    \includegraphics[width=0.4\linewidth]{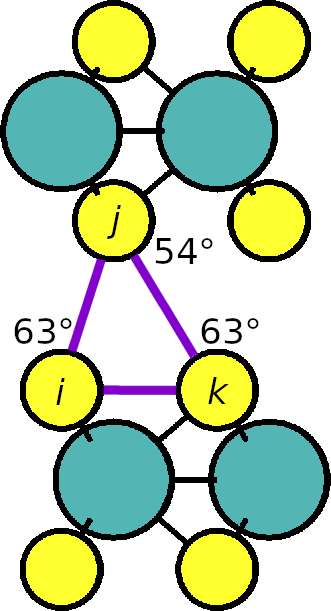}
    \caption{The geometry of a MoS$_2$ bilayer, showing the atomic triples that lead to a large repulsive ATM contribution.}
    \label{fig:mos2}
\end{figure}

More recently, Emrem and coworkers\cite{emrem2022london} benchmarked a variety dispersion corrections for layered materials, including pairwise and non-local schemes. While their results align well with previous studies, they also found that accurate exfoliation energies could be obtained with PBE-D3(0)+ATM.\cite{perdew1996generalized,grimme2010consistent} This method includes the 3-body Axilrod--Teller--Muto (ATM) dispersion term,\cite{axilrod1943interaction,muto1943force} suggesting that dispersion beyond pairwise interactions may be important for layered materials. The ATM term has an angular dependence, with linear configurations of three atoms leading to maximum binding, and atom triples with internal angles of 60$^\circ$ being maximally repulsive. While the ATM term is often negligible for molecular dimers,\cite{otero2013many,otero2020many} von Lilienfeld and Tkatchenko previously noted that molecular dimers containing a $\pi$--$\pi$ stacking interaction have the requisite geometry for large, repulsive three-body interactions.\cite{anatole2010two}  As shown in Figure~\ref{fig:mos2} for the example of MoS$_2$, layered materials have a similar geometry to these stacked dimers, with many ca.\ 60$^\circ$ angles between atomic triples spanning adjacent layers. As such, the LM26 exfoliation energies may be expected to have relatively large ATM contributions, as seen with PBE-D3(0)+ATM. However, the behaviour of the ATM interaction is closely related to its damping function,\cite{emrem2022london,otero2013many} so this finding may or may not be transferable to other dispersion corrections, such as XDM.

XDM has canonically used the Becke--Johnson (BJ) damping function. However, it was recently shown that XDM(BJ) overbinds of clusters of lithium and sodium\cite{becke2022density} due to insufficient damping of the dispersion energy at small internuclear distances for alkali metals. To resolve this, Becke proposed a new damping function with dependence on atomic numbers,\cite{becke2024remarkably} later termed Z damping. XDM(Z) has been shown to give excellent performance for both molecular and solid-state benchmarks, despite having only one functional-dependent empirical parameter.\cite{bryenton2026consistent} However, XDM(Z) has yet to be benchmarked for layered materials, and it remains to be seen where, between the alkali metals and the $p$-block, the performance gap between XDM(BJ) and XDM(Z) closes. Moreover, the importance of the ATM term in XDM for layered materials has not been explored with either BJ or Z damping. 

In this work, we compare the general behaviours of BJ and Z damping for homoatomic dimers spanning all elements from H--Xe. A three-body damping function suitable for XDM(Z) is introduced, and also compared to the three-body BJ-damping function.  Following this, we assess the performance of four XDM variants (BJ/Z damping, with/without ATM) and four D3 variants (BJ/0 damping, with/without ATM) for the LM26 benchmark. 

\section{Theory}\label{sec:theory}

\subsection{Pairwise Dispersion models}

In both the XDM\cite{otero2012van,xdmchapter} and D3\cite{grimme2010consistent,grimme2011effect} corrections, the dispersion energy is written as a sum over all pairs of atoms $i$ and $j$:
\begin{equation}
    E_\text{disp}  = - \sum_{n=6,8,(10)} \, \sum_{i<j} \frac{C_{n,ij} \, f_n(R_{ij})}{R^n _{ij}} \,.
\end{equation}
Here, the $C_{n,ij}$'s are the pairwise dispersion coefficients, $R_{ij}$ is the interatomic distance, and $f_n(R_{ij})$ is a function that damps the dispersion energy to prevent unphysical divergence as $R_{ij}\rightarrow 0$. The D3 correction includes only the $n=6,8$ terms in the dispersion energy, while XDM includes the $n=10$ term as well (although it is typically small). Qualitatively, the $n=6$ term captures instantaneous dipole-dipole interactions between atoms, $n=8$ captures dipole-quadrupole, and $n=10$ captures quadrupole-quadrupole and dipole-octupole interactions. In D3, the $C_{n,ij}$ are computed by interpolating between values for hydride reference compounds based on the atomic coordination numbers.\cite{grimme2010consistent} In contrast, XDM computes $C_{n,ij}$ from  multipole moments derived from the Becke--Roussel exchange hole,\cite{brx} which are functions of the electron density and its derivatives, as well as volume-scaled atomic polarizabilities.\cite{johnson2006post}

The choice of the particular damping function also varies between dispersion corrections. In its original formulation,\cite{grimme2010consistent} D3 uses the Chai--Head-Gordon damping function,\cite{chai2008long} 
\begin{equation}
    f_n ^\text{CHG}(R_{ij}) = \frac{1}{1+6\left( \dfrac{R_{ij}}{s_{r,n}R_{0,ij}}\right)^{-\alpha_n}}\,,
\end{equation}
where $R_{0,ij}$ is an empirical cut-off radius depending on the particular atomic pair, $s_{r,n}$ is a scaling factor, and $\alpha_n$ is a steepness parameter, with $\alpha_6=14$ and $\alpha_8=16$. This function is sometimes referred to as ``zero-damping'' because it prescribes an energy of zero in the united-atom limit.

Both the XDM and D3 dispersion corrections can be paired with the Becke--Johnson damping function,\cite{johnson2006post,grimme2011effect} 
\begin{equation}
    f^\text{BJ} _n(R_{ij}) =\frac{R_{ij} ^n}{R_{ij} ^n + (a_1 R_{c,ij} + a_2)^n}\,,
    \label{e:bjpair}
\end{equation}
where $a_1$ and $a_2$ are empirical parameters, while $R_{c,ij}$ is the ``critical'' radius at which successive terms in the asymptotic expansion of the dispersion energy become equal. As the D3(BJ) correction only includes $C_6$ and $C_8$ terms, $R_{c,ij} = \sqrt{\nicefrac{C_{8,ij}}{C_{6,ij}}}$. In XDM(BJ), $R_{c,ij}$ is the average of three ratios of $C_n$ coefficients: $\sqrt{\nicefrac{C_{8,ij}}{C_{6,ij}}}$, $\sqrt[4]{\nicefrac{C_{10,ij}}{C_{6,ij}}}$, and $\sqrt{\nicefrac{C_{10,ij}}{C_{8,ij}}}$. A key characteristic of the Becke--Johnson damping function is that the damped energy at short range is directly related to the dispersion coefficients. 

XDM can alternatively use the Z-damping function recently proposed by Becke,\cite{becke2024remarkably} 
\begin{equation}
    f_n^\text{Z} (R_{ij}) = \frac{R_{ij}^n}{R_{ij}^n + z_\text{damp}\dfrac{C_{n,ij}}{Z_i + Z_j}}\,,
\end{equation}
where $Z_i$ and $Z_j$ are the atomic numbers of the interacting atom pair, and $z_\text{damp}$ is an empirical parameter. By design, the Z-damping function recovers a correlation energy in the united-atom limit that is unrelated to the $C_n$ coefficients and is equal for each of the $n=6,8,10$ terms:
\begin{equation}
     \lim_{R_{ij}\to0} E_\text{XDM(Z)} = \frac{3(Z_i + Z_j)}{z_\text{damp}}
     \,.
\end{equation}
XDM(Z) gives excellent performance for lattice energies of molecular crystals and for the GMTKN55 benchmark, resolving the overbinding of alkali-metal clusters seen with XDM(BJ).\cite{bryenton2026consistent}

\subsection{3-Body Dispersion Terms}

DFT dispersion corrections may optionally include the Axilrod--Teller--Muto\cite{axilrod1943interaction,muto1943force} (ATM) term, which describes three-body dispersion interactions for an atom triple, $ijk$. The ATM energy expression is
\begin{equation}
    E_{\text{ATM}} = \sum_{i<j<k} \frac{C_{9,ijk}\, f_9
    }{(R_{ij}R_{ik}R_{jk})^3}
    (3\cos\theta_i\cos\theta_j\cos\theta_k +1) \,,
    \label{e:atm_energy}
\end{equation}
where $C_9$ is the ATM dispersion coefficient and $f_9$ is a damping function. Due to the angular dependence, the ATM dispersion energy is attractive for linear arrangements of atoms, but repulsive for interatomic angles near $60^\circ$. 

For the D3(0) dispersion correction, the ATM dispersion coefficient (which is properly a positive quantity) is approximated as the square root of the triple product of the pairwise $C_6$ coefficients for atoms $i$, $j$, and $k$ as
\begin{equation}
    C_{9,ijk}^\text{D3} \approx \sqrt{C_{6,ij}\,C_{6,ik}\,C_{6,jk}} \,.
    \label{e:c9approx}
\end{equation}
While the numerical accuracy of this approximation has been tested, and deviations from the three-body Casimir--Polder integral form\cite{casimir1948influence, stone2013theory} were found to be reasonable,\cite{grimme2010consistent} there is a subtlety regarding the dimensionality of this expression. The $C_n$ coefficients have units of $\text{Ha}\cdot\text{bohr}^n$; thus, the right-hand side of Equation~\ref{e:c9approx} has units of $\text{Ha}^{3/2}\cdot\text{bohr}^9$. For dimensional consistency with the left-hand side, a corrective factor of $\text{Ha}^{-1/2}$ would therefore be required. The $C_9^\text{D3}$ contribution uses a corresponding three-body Chai--Head-Gordon damping function variant,
\begin{equation}
f_9^\text{CHG} =  \frac{1}{1+6\left(\dfrac{\overline{R}_{ijk}}{s_{r,9}\overline{R}_{0,ijk}}\right)^{-16}}\,,
\end{equation}
where $s_{r,9}=\nicefrac{4}{3}$ is a scale factor, $\overline{R}_{ijk}=\sqrt[3]{R_{ij}R_{ik}R_{jk}}$ is the geometric mean of the interatomic distances, and $\overline{R}_{0,ijk}=\sqrt[3]{R_{0,ij}R_{0,ik}R_{0,jk}}$ is the geometric mean of the cutoff radii. The canonical description of the D3(BJ) dispersion correction does not include the ATM term, which was presented only in the context of the earlier D3(0) dispersion model. However, the ATM term is included by default when D3(BJ) is selected in some electronic structure codes, such as Quantum \textsc{Espresso}\cite{giannozzi2017advanced,giannozzi2020quantum} and FHI-aims.\cite{blum2009ab,abbott2025roadmap}

XDM may also include an ATM term with the $C_{9,ijk}$ coefficient computed directly from the squared dipole moment integrals and volume-scaled atomic polarizabilities.\cite{otero2013many} 
The recommended BJ-damping function for the ATM term is a product of three pairwise BJ-damping functions (Equation~\ref{e:bjpair}) of order 3:
\begin{equation}
    f_\text{ATM} ^\text{BJ}= f_3 ^{\text{BJ}}(R_{ij})f_3 ^{\text{BJ}}(R_{ik})f_3 ^{\text{BJ}}(R_{jk}) \,.
\end{equation}
To date, the Z-damping function has not been paired with the ATM term, and an appropriate form using the product of three partial damping functions,
\begin{equation}
    f^\text{Z} _\text{ATM} = f_\text{partial} ^\text{Z}(R_{ij}) f_\text{partial} ^\text{Z}(R_{ik}) f_\text{partial} ^\text{Z}(R_{jk}) \,,
\end{equation}
is proposed here, where
\begin{equation}
    f_\text{partial} ^\text{Z} (R_{ij}) = \frac{R_{ij} ^3}{R_{ij} ^3 + \sqrt{\frac{z_\text{damp} C_{6,ij}}{Z_i + Z_j}}} \,.
\end{equation}
A square-root appears in the denominator to insure an united-atom limit of
\begin{align}\nonumber
    \lim _{\forall R \rightarrow 0} \frac{f^\text{Z} _\text{ATM} C_{9,ijk}}{R_{ij} ^3 R_{ik} ^3 R_{jk} ^3} & = \frac{C_{9,ijk}}{\sqrt{C_{6,ij}C_{6,ik}C_{6,jk}}}\frac{\sqrt{(Z_i + Z_j)(Z_i + Z_k)(Z_j + Z_k)}}{z_\text{damp} ^{3/2}} \\
    & \approx \left( \frac{1}{\sqrt{\text{Ha}}}\right) \frac{\sqrt{(Z_i + Z_j)(Z_i + Z_k)(Z_j + Z_k)}}{z_\text{damp}^{3/2}} \,,
\end{align}
which is effectively independent of the dispersion coefficients. Here, we have leveraged the same approximation as Equation~\ref{e:c9approx} that $C_{9, ijk} \approx \sqrt{{C_{6,ij}C_{6,ik}C_{6,jk}}} \cdot \text{Ha}^{-1/2}$. The corrective factor cancels the $\text{Ha}^{3/2}$ arising from $z_\text{damp}^{-3/2}$ term, ensuring the limit expression has the correct units of energy.

\begin{table*}[ht!]
\caption{Optimal XDM and XDM+ATM BJ-damping ($a_1$, and $a_2$ in \AA) and Z-damping ($z_\text{damp}$, in inverse Hartree) parameters for selected density functionals and basis sets, obtained in this work unless noted otherwise. The mean absolute errors (MAE, in kcal/mol) for the binding energies of the KB49 set of molecular dimers are also shown.}
\label{tab:kb49}
\centering
{\normalsize
\begin{tabular}{lc|ccc|ccc|cc|cc}\hline
Functional & Basis & \multicolumn{3}{c|}{XDM(BJ)}  
                   & \multicolumn{3}{c|}{XDM(BJ)+ATM} 
                   & \multicolumn{2}{c|}{XDM(Z)} &  \multicolumn{2}{c}{XDM(Z)+ATM} \\ 
           &             & $a_1$                               & $a_2$                               & MAE  & $a_1$  & $a_2$  & MAE  & $z_\text{damp}$             & MAE  & $z_\text{damp}$ & MAE \\ \hline
revPBE     & lightdenser & 0.9255$^a$ & 0.3649$^a$ & 0.45 & 0.9805 & 0.1810 & 0.46 & ~39880$^a$ & 0.49 & ~38712 & 0.55 \\
revPBE     & tight       & 0.8992$^a$ & 0.2849$^a$ & 0.36 & 0.9304 & 0.1659 & 0.48 & ~32842$^a$ & 0.42 & ~31749 & 0.47 \\ 
revPBE     & PAW         & 0.4500~~                          & 1.5757~~                          & 0.51 & 0.4133 & 1.6863 & 0.58 & 30160                          & 0.42 & ~29160 & 0.47 \\ \hline
B86bPBE    & lightdenser & 0.6881$^a$ & 1.5789$^a$ & 0.53 & 0.7067 & 1.5069 & 0.55 & 116996$^a$ & 0.60 & 114638 & 0.62 \\
B86bPBE    & tight       & 0.9004$^a$ & 0.7808$^a$ & 0.39 & 0.9040 & 0.7522 & 0.41 & ~96089$^a$ & 0.43 & ~93799 & 0.46 \\ 
B86bPBE    & PAW         & 0.6512$^b$            & 1.4633$^b$            & 0.43 & 0.5781 & 1.6785 & 0.44 & 91449                          & 0.46 & ~89260 & 0.48 \\ \hline
PBE        & lightdenser & 0.3275$^a$ & 2.9627$^a$ & 0.66 & 0.3422 & 2.9050 & 0.68 & 200770$^a$ & 0.74 & 198246 & 0.75 \\
PBE        & tight       & 0.5124$^a$ & 2.2588$^a$ & 0.49 & 0.5103 & 2.2517 & 0.51 & 162373$^a$ & 0.56 & 159797 & 0.58 \\ 
PBE        & PAW         & 0.3275$^b$            & 2.7673$^b$            & 0.51 & 0.2443 & 3.0204 & 0.52 & 153390~~                          & 0.57 & 150899 & 0.59 \\ \hline
\end{tabular}
}

$^a$Values taken from Ref.~\citenum{bryenton2026consistent}.
$^b$Values taken from Ref.~\citenum{xdmchapter}.
\end{table*}

\subsection{Periodic Case} \label{subsec:periodic}

When applied to periodic systems, the pairwise XDM energy expression becomes
\begin{equation}\label{eq:solidpairwise}
    E_\text{XDM} = - \frac{1}{2} \sum_\mathbf{L} \sum_{i\neq j^\prime} \sum _{n=6,8,10} \frac{C_{n,ij} \, f_n(R_{ij})}{R^n _{ij,\mathbf{L}}} \,,
\end{equation}
where $\textbf{L}$ is a displacement lattice vector that accounts for dispersion interactions between atoms in nearby unit cells. The prime excludes the $i=j$ term for lattice vector $\mathbf{L} = 0$.\cite{otero2012van} Similarly, the ATM contribution becomes
\begin{align}\label{eq:solid3body}
    E_\text{ATM} =\frac{1}{6} \sum_{\mathbf{L,M}} \, \sum_{\scriptsize \substack{i \neq j^{\prime} \neq k^{\prime}\\ j\neq k
^{\prime\prime}}} & \frac{C_{9,ijk}\left( 3\cos\theta_i\cos\theta_j\cos\theta_k + 1 \right) }{(R_{ij, \mathbf{L}} ^3) (R_{ik,\mathbf{M}} ^3) (R_{jk, \mathbf{M}-\mathbf{L}} ^3)} \times \notag\\[-1em]
    & \qquad f_\text{ATM}(R_{ij,\mathbf{L}}, R_{ik, \mathbf{M}}, R_{jk, \mathbf{M} - \mathbf{L}}) \,,
\end{align}
where $\mathbf{L}$ and $\mathbf{M}$ are displacement lattice vectors that allow atoms $j$ and $k$ to be in different cells. Again, the prime excludes the $i=j$ and $i=k$ term for lattice vector $\mathbf{L} = 0$ and the double prime excludes the $j=k$ term for all lattice vectors $\mathbf{L}=\mathbf{M}$. 

For both atomic pairs and triples, convergence is achieved when the change in dispersion energy from expanding the sums over cells in Eqns.~\ref{eq:solidpairwise} and \ref{eq:solid3body} to include successive periodic images is below $1\times10^{-6}$~eV. To speed up evaluation of the dispersion energy for large supercells, a radial cutoff of
\begin{equation}
    r_\text{max} = \left( \frac{\max\{C_{6,ij}\}}{10^{-12}~\text{Ha}} \right) ^{1/6}
\end{equation}
is imposed, such that atom pairs separated by distances greater than $r_\text{max}$ do not contribute. 
Because of the sum over two sets of periodic images in the ATM case, direct convergence can be very slow. However, convergence with respect to the number of image cells is approximately log-linear. We propose a scheme to extrapolate the converged energy as a geometric series. During ATM convergence, when the last seven changes in energy are log-linear with correlation coefficient $r^2>0.9925$, the converged energy is estimated as 
\begin{equation}
    E_\text{conv} = E_\text{unconv} +  \frac{s}{1-s} \Delta E \,,
\end{equation}
where $s$ is the slope of the log-linear energy convergence, 
describing the average change in energy with each additional periodic image included in the summation. Because deviations from log-linearity will occur at near-converged energies, the errors of the extrapolation compared to direct convergence are very small ($\ll1$\%). A visual comparison of the direct convergence and the extrapolation method can be found in the ESI. If the energy converges to within $1\times10^{-6}$~eV before log-linearity is achieved, the direct energy convergence is used. 

\section{Computational Methods}\label{sec:methods}

\subsection{Free Atoms}

To investigate differences between BJ and Z damping, calculations were performed for all isolated atoms for elements ranging from H--Xe using their known ground-state electronic configurations. Single-point energies were evaluated with PBE0\cite{adamo1999toward}/def2-TZVP\cite{weigend2005balanced} using the Gaussian16 program.\cite{frisch2016gaussian} For atoms with unpaired d electrons, the d orbitals were populated according to the prescription given in Ref.~\citenum{johnson2007density}, with two unpaired d electrons or holes occupying the d$_{z^2}$ and d$_{x^2-y^2}$ orbitals (the $e_g$ orbitals in the octahedral point group). The resulting electron densities were saved as \texttt{.wfx} files and used as input to the \texttt{postg} program,\cite{otero2025postgxcdm} which was used to compute the XDM dispersion coefficients.

\begin{figure*}[t!]
    \includegraphics[width=\textwidth]{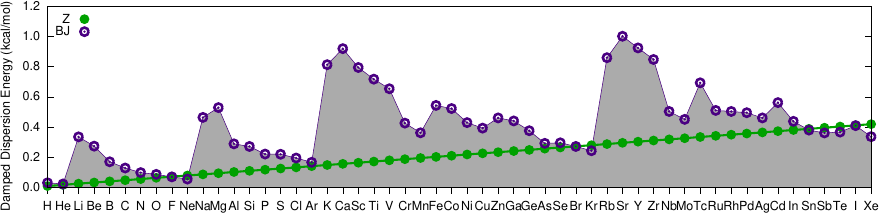}
    \caption{Comparison of BJ- and Z-damping functions for the elements H--Xe. Results are the damped dispersion energies for a homoatomic dimer in the united-atom (i.e.\ $R\rightarrow0$) limit including only the leading-order $C_6$ term. The plot uses the XDM dispersion coefficients computed for the free atoms with PBE0/def2-TZVP using Gaussian 16, together with the FHI-aims \texttt{tight} damping parameters.}
    \label{fig:damping}
\end{figure*}

Figure~\ref{fig:damping} shows the BJ- and Z-damped energies in the united-atom ($R\rightarrow0$) limit for the homoatomic dimers consisting of the elements ranging from H--Xe.  By design, the XDM(Z) energy in the united-atom limit scales linearly with the atomic number and is independent of the electronic configuration. On the other hand, the XDM(BJ) results depend on the electron density and, consequently can give different united-atom limits for the same element depending on its spin state, as shown in the ESI for the examples of nickel and palladium. The XDM(BJ) united-atom energies follow the trend expected for atomic radii, increasing down a group of the periodic table while decreasing across a period. It is this behaviour that results in strong overbinding for alkali-metal clusters,\cite{becke2022density,becke2024remarkably} and similar errors appear likely for other $s$-block metals like calcium and strontium, as well as early-row transition metals like scandium, yttrium, titanium, and zirconium. The XDM(BJ) and XDM(Z) energies achieve better alignment later in a given period, with notable agreement for the halogens and noble gases in the united-atom limit.

\subsection{The KB49 Benchmark and XDM Parametrisation}

XDM(BJ) has long been available for use with GGA functionals, planewave (PW) basis sets, and projector-augmented wave (PAW) pseudopotentials in the Quantum \textsc{Espresso} program.\cite{otero2012van,giannozzi2017advanced,giannozzi2020quantum} Recently,\cite{bryenton2026consistent} both XDM(BJ) and XDM(Z) were parametrised for use with 16 density functionals and 5 numerical atomic orbital (NAO) basis sets using the FHI-aims program.\cite{blum2009ab, yu2018elsi, havu2009efficient, ihrig2015accurate} In the present work, damping parameters are fit to allow use of XDM(Z) with PAW calculations using Quantum \textsc{Espresso} for three GGA functionals: B86bPBE,\cite{becke1986large,perdew1996generalized} PBE,\cite{perdew1996generalized} and revPBE.\cite{zhang1998comment} Additionally, both BJ- and Z-damping parameters were reoptimised for use in conjunction with the ATM term for the same three functionals. A similar XDM reparametrisation was also performed for the lightdenser and tight NAO basis sets using FHI-aims. 

The optimal damping parameters were determined by minimising the root-mean-square percent error (RMSPE) for the KB49 molecular benchmark.\cite{kannemann2010van} Geometries and reference binding energies for the 49 molecular complexes are available in the \texttt{refdata} repository.\cite{otero2015refdata} The Quantum \textsc{Espresso} calculations used kinetic energy and density cutoffs of 80.0 and 800.0~Ry, respectively, and were performed at the $\Gamma$-point. The FHI-aims calculations used the ZORA relativistic correction for all elements.\cite{van1994relativistic} 
ATM calculations were performed using the external \texttt{parrotXDM}\cite{parrotxdm} tool, which implements the extrapolation method described in Section~\ref{subsec:periodic}. The \texttt{parrotXDM} code reads the atomic positions, multipole moments, and atomic polarisabilities from Quantum \textsc{Espresso} or FHI-aims output files to compute the 3-body XDM dispersion energy..

A summary of the final XDM damping parameters, as well as the corresponding KB49 error statistics can be found in Table~\ref{tab:kb49}. The single $z_\text{damp}$ parameter gives clear insight into the extent of damping for a given functional. Similar trends can be observed for BJ damping, but they are obscured by the linear dependence of the $a_1$ and $a_2$ parameters.\cite{otero2012van} As revPBE is the least attractive functional considered, it has the lowest $z_\text{damp}$ parameter, indicating weak damping. PBE, being the most attractive, gives the largest $z_\text{damp}$ parameters. Increasing the basis-set size for a given functional from lightdenser, to tight, to planewaves, generally decreases the magnitude of the damping parameters. This occurs because smaller basis sets are more likely to lead to basis-set superposition errors; this causes an increase in binding energy and requires greater damping of the dispersion term to offset the error. Finally, as the ATM term is weakly repulsive for most complexes, including it causes a small decrease in the damping parameters, and its inclusion serves to worsen the KB49 benchmark statistics slightly. 

\subsection{The LM26 benchmark}

The LM26 benchmark consists of reference RPA exfoliation energies and $c$ lattice parameters for 23 transition metal dichalcogenides, as well as lead(II) oxide, graphite, and hexagonal boron nitride.\cite{bjorkman2014testing}
To compare with the reference data, the required DFT calculations consist of a sequence of single-point energy evaluations with the $c$-parameter and interlayer separation gradually varied to generate a potential energy curve. 
A denser sampling of points is used around the expected minimum, and a coarser sampling near the fully exfoliated end point (with 15 \AA\ of added interlayer space). The minimum energy and corresponding $c$ parameter are found by interpolation. The exfoliation energy is then taken as the energy difference between the exfoliated end point and the minimum energy obtained from the interpolation.

Calculations were performed using either the Quantum \textsc{Espresso}\cite{giannozzi2017advanced,giannozzi2020quantum} or FHI-aims\cite{blum2009ab, yu2018elsi, havu2009efficient, ihrig2015accurate, price2023xdm,bryenton2026consistent} programs. The Quantum \textsc{Espresso} settings were the same as those used in Ref.~\citenum{otero2020asymptotic}, consisting of \texttt{ecutwfc} and \texttt{ecutrho} values of 100.0 and 1000.0~Ry, respectively, and 12$\times$12$\times$4 \textbf{k}-grids. Cold smearing\cite{marzari1999thermal} was applied to aid convergence for some materials, with widths of 0.01~Ry for \ch{NbTe2} and 0.001~Ry for graphite, \ch{TaS2}, \ch{TaSe2}, \ch{VS2}, and \ch{VSe2}. PAW datasets were obtained from Dal Corso's \texttt{pslibrary}.\cite{corso2014pseudopotentials} FHI-aims calculations were performed using version 260302, as this version outputs the atomic volumes and multipole moments used by \texttt{parrotXDM}\cite{parrotxdm} to compute three-body contributions to the dispersion energy. Calculations used either the lightdenser\cite{bryenton2026consistent} or tight basis settings, and the ZORA relativistic correction\cite{van1994relativistic} for all elements.

Dispersion effects were described with four variants of each of the XDM\cite{otero2012van,xdmchapter} and D3\cite{grimme2010consistent,grimme2011effect} methods, depending on the choice of damping function and inclusion or exclusion of the ATM term. The XDM calculations used the  B86bPBE,\cite{becke1986large} PBE,\cite{perdew1996generalized} and revPBE\cite{zhang1998comment} functionals, while the D3 calculations only considered PBE and revPBE due to a lack of available D3 damping parameters for use with B86bPBE. Additional calculations were also performed using the LSDA (Perdew--Zunger\cite{perdew1981self} for Quantum \textsc{Espresso}, Perdew--Wang\cite{perdew1992accurate} for FHI-aims) and PBEsol,\cite{perdew2008restoring} without any dispersion correction.

\section{Results and Discussion}




\begin{table*}[ht!]
    \centering
    \caption{Mean absolute errors (MAE) and mean errors (ME) in meV/\AA$^2$ for the LM26 benchmark of exfoliation energies of 26 layered materials, including graphite, boron nitride, lead(II) oxide, and transition-metal dichalcogenides. A positive ME indicates overbinding.}
    \setlength{\tabcolsep}{6pt}
    \begin{tabular}{l|rr|rr|rr}
        \hline
        Functional        
        & \multicolumn{2}{c|}{lightdenser} & \multicolumn{2}{c|}{tight} & \multicolumn{2}{c}{PW} \\ 
                              &  MAE &   ME  &  MAE &   ME  &  MAE &   ME  \\ \hline
        revPBE-D3(BJ)         & 35.1 &  35.1 & 35.0 &  35.0 & 34.5 &  34.5 \\
        revPBE-D3(BJ)+ATM     & 27.8 &  27.8 & 27.7 &  27.0 & 27.2 &  27.2 \\
        revPBE-D3(0)          & 16.5 &  16.5 & 16.2 &  16.2 & 15.8 &  15.8 \\
        revPBE-D3(0)+ATM      & 10.2 &  10.0 &  9.9 &   9.6 &  9.6 &   9.2 \\
        PBE-D3(BJ)            & 14.7 &  14.6 & 14.5 &  14.3 & 12.8 &  12.6 \\
        PBE-D3(BJ)+ATM        &  8.2 &   7.8 &  8.1 &   7.5 &  6.9 &   6.1 \\  
        PBE-D3(0)             &  9.8 &   9.6 &  9.5 &   9.2 &  8.4 &   8.0 \\
        PBE-D3(0)+ATM         &  4.1 &   3.2 &  4.0 &   2.8 &  3.7 &   1.8 \\  \hline
        revPBE-XDM(BJ)        &  5.0 &   5.0 & 11.5 &  11.5 &  6.4 &   4.7 \\
        revPBE-XDM(Z)         &  6.5 &   6.1 &  8.2 &   7.8 &  5.4 &   4.2 \\
        revPBE-XDM(BJ)+ATM    &  2.2 &  -0.9 &  5.2 &   5.2 &  4.4 &  -2.2 \\
        revPBE-XDM(Z)+ATM     &  4.6 &  -0.3 &  5.3 &   0.8 &  5.3 &  -2.0 \\
        B86bPBE-XDM(BJ)       &  5.9 &   5.9 &  7.7 &   7.7 &  4.7 &   3.3 \\
        B86bPBE-XDM(Z)        &  5.8 &   5.5 &  7.3 &   7.1 &  5.3 &   4.5 \\
        B86bPBE-XDM(BJ)+ATM   &  2.3 &   1.2 &  3.5 &   3.2 &  3.1 &  -0.2 \\
        B86bPBE-XDM(Z)+ATM    &  3.0 &   1.3 &  4.0 &   2.6 &  3.5 &   0.4 \\
        PBE-XDM(BJ)           &  5.6 &   5.5 &  6.8 &   6.7 &  4.4 &   1.6 \\ 
        PBE-XDM(Z)            &  4.1 &   3.2 &  5.4 &   4.8 &  4.6 &   1.8 \\
        PBE-XDM(BJ)+ATM       &  2.7 &   0.9 &  3.3 &   2.1 &  3.9 &  -2.2 \\ 
        PBE-XDM(Z)+ATM        &  2.6 &  -0.1 &  3.2 &   1.0 &  3.9 &  -1.4 \\ \hline
        revPBE                & 19.8 & -19.8 & 20.1 & -20.1 & 20.3 & -20.3 \\
        B86bPBE               & 18.6 & -18.6 & 19.2 & -19.2 & 19.5 & -19.5 \\
        PBE                   & 17.7 & -17.7 & 18.3 & -18.3 & 18.6 & -18.6 \\
        PBEsol                & 11.9 & -11.9 & 12.1 & -12.1 & 12.7 & -12.7 \\ 
        LDA                   &  3.7 &  -1.9 &  4.0 &  -2.2 &  4.0 &  -2.4 \\ \hline
    \end{tabular}
    \label{tab:maes}
\end{table*}

\begin{figure}[ht]
    \centering
    \includegraphics[width=\linewidth]{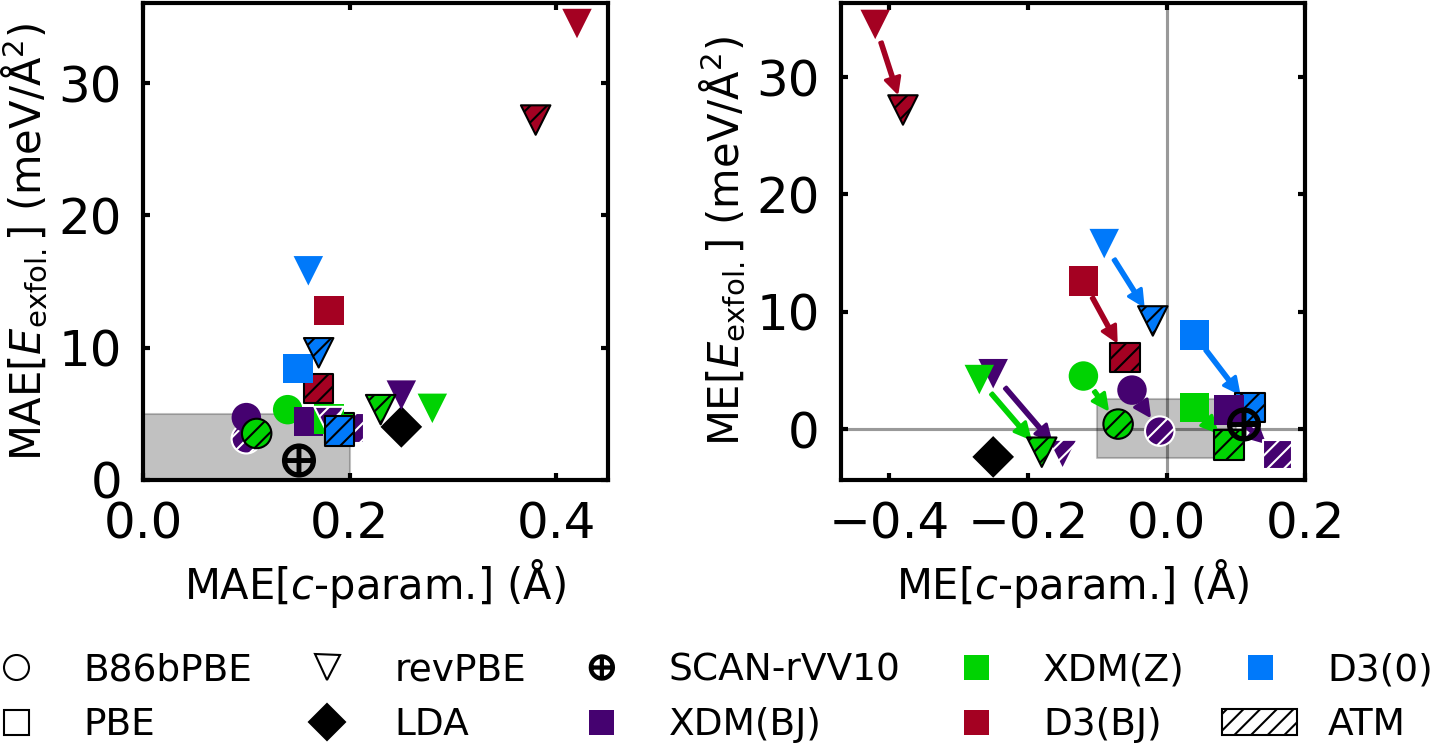}
    \caption{Plot of the exfoliation-energy and $c$-parameter MAEs and MEs with PW/PAW with selected functionals and dispersion corrections. Note that results for dispersion-uncorrected GGAs are not shown as the errors exceed the bounds of the plots. The grey box highlights acceptable margins of error: 5~meV/\AA$^2$ and 0.2~\AA, although we note that the RPA reference data may underestimate the true exfoliation energies by roughly this amount. The SCAN-rVV10 error for LM26 is also shown for comparison.\cite{ning2022workhorse} Arrows annotate the mean error plot (right) to show the trend in the energy and $c$ parameter values with the inclusion of the ATM term.}
    \label{fig:paw_exfol_c}
\end{figure}

Full tables of the computed exfoliation energies and $c$ lattice parameters for the LM26 benchmark are included in the ESI. Table~\ref{tab:maes} summarises the exfoliation-energy results; it contains the mean absolute errors (MAE) and mean errors (ME) for all functional/basis/dispersion correction combinations considered. An analogous table for the MAEs and MEs in $c$ lattice parameters is also given in the ESI. Finally, Figure~\ref{fig:paw_exfol_c} shows scatter plots of the exfoliation-energy errors versus the lattice-parameter errors, for the case of the planewave/PAW calculations only, as these are most reflective of the basis-set limit. Analogous plots using the NAO-basis data are, again, shown in the ESI.

Beginning with the dispersion-uncorrected GGA functionals, all are shown to underbind severely, as expected \cite{rydberg2003van,bjorkman2012we} from the lack of dispersion. While the LDA gives reasonably accurate exfoliation energies, it underestimates the $c$ lattice parameters, as shown in Figure~\ref{fig:paw_exfol_c}. The good performance for exfoliation energies is the result of fortuitous error cancellation between the overbinding nature of the LDA's exchange functional and its neglect of London dispersion physics. Overall, the ordering of the exfoliation energy errors in Table~\ref{tab:maes} (viz.\ revPBE $>$ B86b $>$ PBE $>$ PBEsol $>$ LDA) matches the ordering of the exchange enhancement factors of these functionals for moderate to large reduced density gradients.\cite{price2021requirements}

The D3 results obtained here are also consistent with previous works.\cite{tawfik2018evaluation,otero2020asymptotic,emrem2022london} GGA functionals paired with any of the D3 corrections give exfoliation energies that are consistently too large in magnitude, indicating overbinding. It was argued previously that this was due to overestimation of the dispersion coefficients by the D3 method.\cite{otero2020asymptotic} Only PBE-D3(0)+ATM gives exfoliation energies approaching the target accuracy of errors $<$ ca.\ 5 meV/\AA$^2$,\cite{otero2020asymptotic} as noted by Emrem \textit{et al.}\cite{emrem2022london} However, for prediction of the $c$ lattice parameters, all D3-corrected functionals perform reasonably well, except for revPBE-D3(BJ) with or without ATM terms.  These results align with previous literature,\cite{tawfik2018evaluation,otero2020asymptotic,emrem2022london} which found that asymptotic dispersion schemes are generally accurate for geometric quantities.

Turning to the XDM-based methods, they tend to be more consistently accurate for the exfoliation energies than most D3-based methods, aligning with our previous findings.\cite{otero2020asymptotic} However, as for the D3-based methods, XDM-based methods also give larger errors in both exfoliation energies and lattice parameters when paired with the revPBE functional. While revPBE (and its hybrid counterpart, revPBE0) shows excellent performance for the GMTKN55 molecular benchmark,\cite{goerigk2017look} its performance compared to PBE/B86bPBE deteriorates for molecular crystals,\cite{bryenton2026consistent} and we see here that its performance deteriorates further for layered materials. Thus, we do not recommend the revPBE functional for use in the solid state due to its inconsistent performance.

We also consider the effect of the ATM term; its minimum, maximum, and average contributions to the exfoliation energy can be found in the ESI. As expected from the prevalence of near-equilateral-triangle arrangements of neighbouring atoms in layered materials (shown in Figure~\ref{fig:mos2}), the ATM term is typically repulsive, destabilizing the bulk and resulting in smaller exfoliation energies. The annotations in Figure~\ref{fig:paw_exfol_c} clearly show this repulsive, destabilizing behaviour, where the ME for the exfoliation lowers and that for the $c$ parameter slightly rises. The only two exceptions occurred for PbO with PBE-XDM(BJ)+ATM/PW and h-BN with revPBE-XDM(Z)+ATM/PW, for which the ATM term is very weakly attractive. Following the strength of the damping parameters, the magnitude of the XDM+ATM term is largest for revPBE (least damped), and smallest for PBE (most damped). The magnitude of the ATM term is also somewhat larger for D3(0)+ATM than for XDM(BJ) or XDM(Z).

For molecular clusters, dimers, and crystals, the ATM contribution is often small;\cite{otero2020many} however, it has been shown to be appreciable for large molecular systems,\cite{distasio2012collective,anatole2010two} and also for the LM26 in its PBE-D3(0)+ATM parametrisation.\cite{emrem2022london} From the results in Table~\ref{tab:maes}, the ATM term is significant for the LM26 when used in conjunction with XDM as well. XDM(BJ) and XDM(Z) tend to overbind (by ca.\ 2-8 meV/\AA$^2$ depending on basis set when paired with B86bPBE and PBE), although this should be taken with a grain of salt due to the previously noted underbinding tendency of the RPA.\cite{krogel2020perspectives, ning2022workhorse} Addition of the ATM term 
brings XDM into exceptional agreement with the RPA for exfoliation energies; it  routinely improves the MAE, and improves the ME in most cases, although it occasionally overcompensates and results in a small consistent underbinding.  The ATM term also causes a slight lengthening of the predicted $c$ lattice parameters by 
0.06~\AA\ or 0.8\% on average, tending to improve accuracy when paired with B86bPBE, but not with PBE.

Comparing the damping functions, XDM(Z) tends to be as accurate as XDM(BJ), and the average MAEs in the exfoliation energies across all functionals and basis sets are 4.9~meV/\AA$^2$ with both damping functions. Using the planewave basis, which is approaching the complete basis-set limit, Figure~\ref{fig:paw_exfol_c} shows that the best simultaneous accuracy for exfoliation energies and lattice parameters for LM26 is obtained with B86bPBE-XDM(BJ)+ATM and B86bPBE-XDM(Z)+ATM. These, along with several other XDM-corrected functionals, provide a level of accuracy approaching that of the more computationally demanding non-local SCAN-rVV10 method for the exfoliation energies. Further, the semi-local methods presented here rival or outperform SCAN-rVV10 for the prediction of the $c$ lattice parameter.
However, all of the errors are measured with respect to the RPA, which is expected to underbind, perhaps by as much as 5--10 meV/\AA$^2$ based on what limited QMC data is available.\cite{krogel2020perspectives, ning2022workhorse} Thus, the performance of SCAN-rVV10 and all combinations of B86bPBE and PBE with the various XDM variants is effectively the same, as all are accurate to within the RPA error bars. This highlights the need for higher level reference data for the LM26 benchmark, as dispersion-corrected density functionals have now outgrown comparison to the RPA.

Finally, the results in Table~\ref{tab:maes} show that the lightdenser NAO basis enjoys some error cancellation from basis-set superposition error, which results in the smallest exfoliation-energy errors yet reported for a GGA paired with an asymptotic dispersion correction for the LM26 benchmark. The lower computational cost means that the lightdenser basis set would be recommended for calculations on large systems. Lastly, the effect of the ATM term on the $c$ parameters is small relative to its contribution to the exfoliation energy. It may, therefore, be considered as a refinement in a final single-point energy calculation to avoid the computational expense of summing over atomic triples during geometry optimisations.

\section{Summary}

This work explored the influence of the damping function and the inclusion of the $C_9$ (ATM) term on the DFT description of the exfoliation energies and out-of-plane lattice constants of twenty-six layered materials. While GGA functionals always underestimate exfoliation energies, the addition of a pairwise dispersion scheme can overcompensate, leading to a systematic overbinding tendency. This is quite evident for the D3 correction, where BJ damping overbinds more than zero-damping. The extent of the overbinding is typically reduced with the XDM dispersion correction, for which BJ and Z damping give similar accuracy on average. However, we recommend the Z damping function in general because of its greater simplicity, lesser empiricism, and more consistent energies in the united-atom limit. 

We further demonstrated that the three-atom ATM term is relatively large and repulsive for layered materials. Thus, its inclusion reduces the residual overbinding found with all pairwise-only dispersion methods, lessening the errors relative to RPA reference data. Inclusion of the ATM term in XDM leads to the lowest errors yet reported for the LM26 benchmark using an asymptotic dispersion correction. In the basis-set limit, B86bPBE-XDM(BJ)+ATM and B86bPBE-XDM(Z)+ATM are the top-performing semi-local functionals considered here, while error cancellation leads to slightly improved statistics with the lightdenser basis set.
Future work will consider implementing the ATM correction to both XDM(BJ) and XDM(Z) directly in FHI-aims. 

However, despite the lower errors for the LM26 benchmark offered by the ATM term, we do not recommend it as a universal, default component in XDM due to its poor computational scaling for periodic systems and generally small energy contributions for molecular complexes and crystals. Even for layered materials, the impact of the ATM term on the crystal structure is minor, resulting in only a 0.06~\AA\ increase in the out-of-plane ($c$-axis) lattice parameters on average. This means that the ATM term may only be included as a final energy correction and not as part of a geometry optimisation.
In practice, we would recommend either variant of B86bPBE-XDM with the lightdenser settings for geometry optimisations, followed by single-point energy evaluation with the ATM term added, as the best balance of computational cost and accuracy. However, there is little difference between the B86b and PBE exchange functionals for these layered materials, particularly given the uncertainty of the reference data, and this choice can be left to the preference of the user. 
Finally, we note that the RPA reference exfoliation energies now have large uncertainties relative to the level of accuracy that can be achieved with the best dispersion-corrected density functionals. It is, therefore, clear that there is an ongoing need for more accurate reference data for exfoliation energies of layered materials.

\section*{Acknowledgements}

AFR, KRB, and ERJ thank the Natural Sciences and Engineering Research Council (NSERC) of Canada for financial support and the Atlantic Computing Excellence Network (ACENET) for computational resources. ERJ additionally thanks the Royal Society for a Wolfson Visiting Fellowship.
    
\section*{Data Availability Statement}

The data that support the findings of this study are available in the supplementary information.

\section*{Conflicts of Interest}

There are no conflicts to report.

\balance

\bibliographystyle{rsc}
\bibliography{main.bib}

\end{document}